\documentclass[aps,prl,twocolumn,amsmath,amssymb,showpacs,superscriptaddress,notitlepage]{revtex4-1}
\usepackage[T1]{fontenc}
\usepackage{color}
\usepackage{bm}
\usepackage{amsmath}
\usepackage{amssymb}
\usepackage{graphicx}
\usepackage{esint}
\usepackage{mathrsfs}
\usepackage{booktabs}
\usepackage{multirow}
\usepackage{makecell}
\usepackage{verbatim}
\usepackage[plainpages = false, pdfpagelabels, 
                 bookmarks,
                 bookmarksopen = true,
                 bookmarksnumbered = true,
                 breaklinks = true,
                 linktocpage,
                 colorlinks = true,
                 linkcolor = blue,
                 urlcolor  = blue,
                 citecolor = blue,
                 anchorcolor = green,
                 hyperindex = true,
                 hyperfigures
                 ]{hyperref} 
\usepackage{dcolumn}   

\makeatletter
\newcommand*{\rom}[1]{\expandafter\@slowromancap\romannumeral #1@}
\makeatother

\usepackage{times}{\Large }

\makeatother

\usepackage{babel}

\begin{document}

\title{Influence of topological degeneracy on the boundary Berezinskii-Kosterlitz-Thouless quantum phase transition of a dissipative resonant level}

\author{Gu Zhang}
\email{zhanggu@baqis.ac.cn}
\affiliation{Beijing Academy of Quantum Information Sciences, Beijing 100193, China}

\author{Zhan Cao}
\affiliation{Beijing Academy of Quantum Information Sciences, Beijing 100193, China}

\author{Dong E. Liu}
\email{dongeliu@mail.tsinghua.edu.cn}
\affiliation{State Key Laboratory of Low Dimensional Quantum Physics, Department of Physics, Tsinghua University, Beijing, 100084, China}
\affiliation{Beijing Academy of Quantum Information Sciences, Beijing 100193, China}
\affiliation{Frontier Science Center for Quantum Information, Beijing 100184, China}
\affiliation{Hefei National Laboratory, Hefei 230088, China}

\date{\today}

\begin{abstract}
The interplay between a topological degeneracy and the residue degeneracy (also known as the residue entropy) of quantum criticality remains as an important but not thoroughly understood topic.
We find that this topological degeneracy,
provided by a Majorana zero mode pair, relaxes the otherwise strictly requested symmetry requirement, to observe the boundary Berezinskii-Kosterlitz-Thouless (BKT) quantum phase transition (QPT) of a dissipative resonant level.
Our work indicates that the topological
degeneracy can be potentially viewed as an auxiliary symmetry that realizes a robust boundary QPT.
The relaxation of the symmetry requirement extends the transition from a point to a finite area, thus greatly reducing the difficulty to experimentally observe the QPT.
This topology-involved exotic BKT phase diagram, on the other hand, provides another piece of evidence that can further confirm the existence of a Majorana zero mode.
\end{abstract}

\maketitle


\textbf{\emph{Introduction}}---Boundary quantum phase transitions (QPTs)~\cite{VojtaBQPT,AffleckQuamtumImpurityX} are special QPTs triggered by the change of boundary conditions.
Similar as the bulk ones, boundary QPTs normally involve symmetries.
For instance, in multi-channel Kondo~\cite{AffleckLudwigNPB91,AffleckLudwigPRB92,EmeryKivelsonPRB92,MitchellSelaPRL16,LopesAffleckSelaPRB20}, Luttinger liquid resonant~\cite{KomnikGogolinPRL03}, multiple-impurity Anderson or Kondo model~\cite{JonesVarmaPRL87,JonesKotliarMillisPRB89,AffleckLudwigPRL92,KrishnamurthyPRB93,GarstVojtaPRB04,ZitkoRPB06,ZitkoPRL12}, and dissipative resonant level~\cite{Mebrahtu12,Mebrahtu13,DongHuaixiuPRB14} models, the observation of quantum critical point (QCP) requires the impurity to symmetrically couple to involved leads.
Actually, the symmetric coupling that triggers quantum frustration at QCP produces a non-trivial residue degeneracy, which is another important feature of QPT~\cite{AffleckLudwigNPB91,VojtaBQPT}.
This residue degeneracy, heavily focused on recently as a QPT-induced ``anyon''~\cite{SelaPRB19,SelaPRB20,YasharPRB20,GoldsteinPRL22,SelaPRB22}, has been proposed as an amplifier of topological non-triviality~\cite{ChristianPRB20}.
As another example, a Berezinskii-Kosterlitz-Thouless (BKT) transition is predicted in a dissipative resonant level model (DRLM)~\cite{ChunghouPRL09,DongHuaixiuPRB14}, between the ballistic-transmission and isolated-impurity phases.
Its observation, importantly, also requires perfectly symmetric lead-dot couplings~\cite{ChunghouPRL09,DongHuaixiuPRB14}.
This fine-tuning prerequisite of symmetry is among the difficulties to observe this BKT QPT.
To the best of our knowledge, a BKT-type boundary QPT, has not yet been reported experimentally.

On the other hand, a pair of Majorana zero modes (MZMs), predicted to exist at end points of a nanowire hetero-structure~\cite{Kitaev2001UFN,Lutchyn2010PRL,Oreg2010PRL} and vortices of a topological superconductor~\cite{Read-2000-PRB,Fu&Kane-MF}, is known to provide a non-locally defined topological degeneracy, given a long enough inter-MZM distance.
In addition, the coupling between a topological degeneracy and a lead always has equal electron (i.e., normal tunneling) and hole (i.e., Andreev reflection) components.
These two robust features of a topological degeneracy, distinct from a trivially local one, are among the key elements of topological quantum computation~\cite{Nayak2008RMP}.
Exploring the intricate influences of topological degeneracy on the residue degeneracy in QPTs, where symmetry is a prerequisite, constitutes a compelling avenue of investigation. A crucial line of inquiry involves investigating the potential role of topological degeneracy as an auxiliary symmetry capable of reinstating boundary QPTs that have been undermined by asymmetry.

\begin{figure}
  \includegraphics[width=\linewidth]{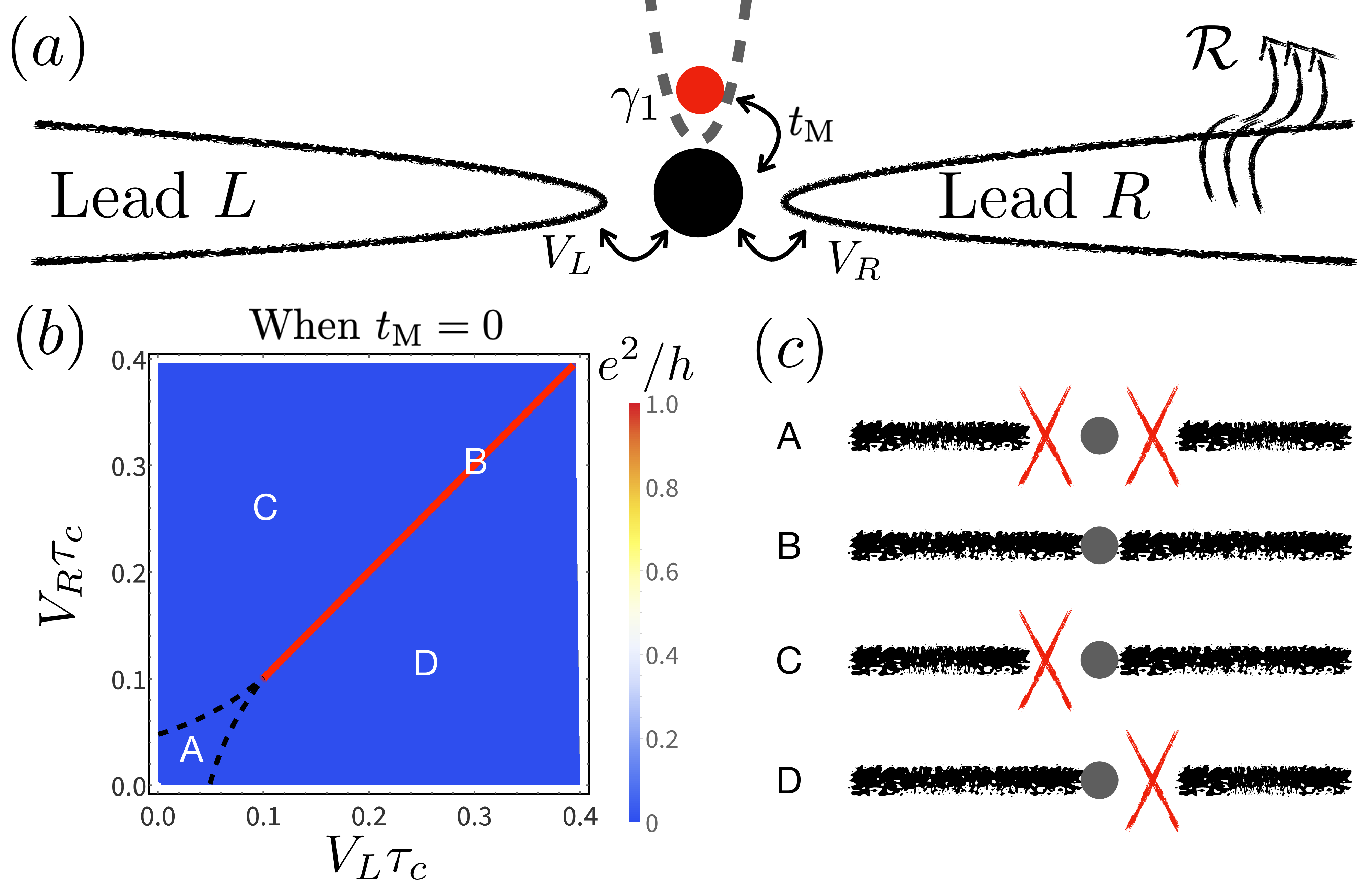}
  \caption{(a) The model of the system.
  The MZM (the red dot) couples to the dot of the DRLM, with the amplitude $t_\text{M}$. The dot couples to two leads with corresponding amplitudes $V_L$ and $V_R $.
  The leads are dissipative, with the \textit{total} impedance $\mathcal{R}$. When $t_\text{M} = 0$, the system reduces to the DRLM, with the zero-temperature equilibrium conductance in (b) (when $r = 2.5$), as a function of dimensionless quantities $V_L \tau_c$ and $V_R \tau_c$. Here $\tau_c$ is the cutoff in imaginary time.
  In real systems, $\tau_c \sim 1/D$, with $D$ the relevant bandwidth~\cite{SchillerIngersentRG97}. 
  In (b), the conductance is determined by the fixed points [with four candidates in (c)] to which the system approaches [figured out by RG equations Eqs.~\eqref{eq:high-t-flow} and \eqref{eq:low-t-flows}] at zero temperature.
  The marked-out phases A, B, C, and D are illustrated in (c), with red crosses indicating vanishing lead-dot communications.
  Of phase B, both dot-lead couplings are transparent, leading to a perfect equilibrium conductance $e^2/h$, at zero temperature. Otherwise (phases A, C and D), dot decouples with one or both leads. The corresponding zero-temperature equilibrium conductance vanishes.
  }
  \label{figure:model}
\end{figure}

\begin{figure}
  \includegraphics[width=\linewidth]{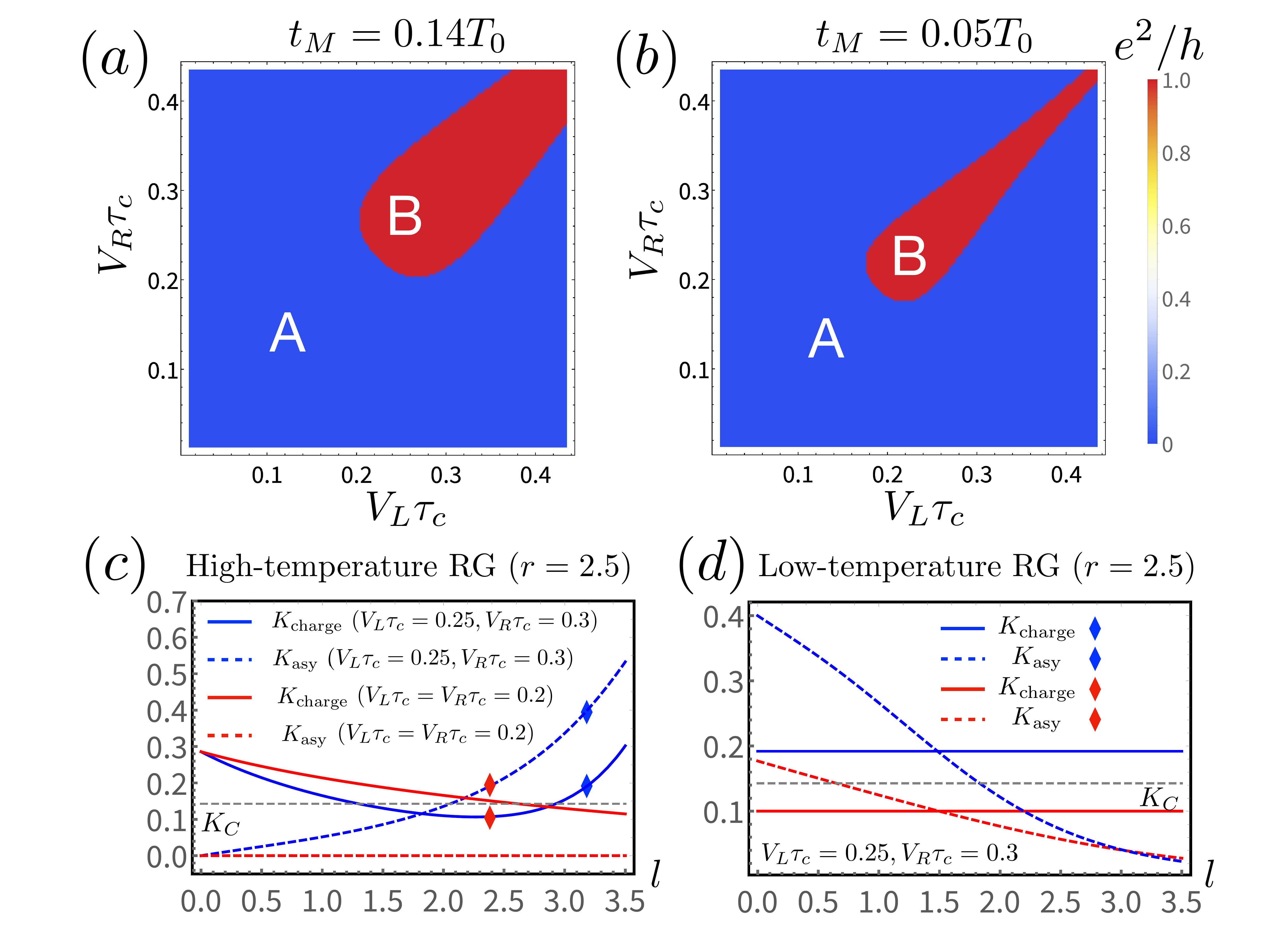}
  \caption{
  Phase diagrams (indicated by the corresponding zero-temperature equilibrium conductance) and RG flow curves when $r = 2.5$. 
  (a), (b) are the phase diagrams with larger ($t_\text{M} = 0.14 T_0$) and smaller ($t_\text{M} = 0.05 T_0$) values of $t_\text{M}$, respsctively.
  Here $T_0$ is the initial temperature of RG flow (i.e., the temperature at which parameters in Hamiltonian are obtained) that requires experimental calibration.
  Similar as that in Fig.~\ref{figure:model}(b), the conductance is obtained after identifying the fixed point [A or B, cf. Fig.~\ref{figure:model}(c)] to which the system flows, at zero temperature.
  (c) High-temperature RG flows curves of $K_\text{charge}$ and $K_\text{asy}$, for the symmetric (red curves) and asymmetric (blue curves) situations.
  The red and blue diamonds marked out points that belong to the ballistic-transmission ($K_\text{charge} < K_C$) and isolated-impurity ($K_\text{charge} > K_C$) phases, respectively, if high-temperature RG flow terminates at their corresponding $l$ ($l=0$ when RG initially begins).
  (d) Low-temperature RG flow of $K_\text{charge}$ and $K_\text{asy}$, with their initial values given by the corresponding diamond.
  }
  \label{fig:phase_diagrams}
\end{figure}

In this work, we thus investigate the influence of
topological
degeneracy on the boundary BKT QPT in a DRLM.
Without the topological
degeneracy, it is known that the BKT transition of a DRLM requires a perfect symmetry between two involved leads.
Surprisingly, as among our central results, this strict symmetry requirement (to observe the BKT QPT) is remarkably relaxed, by coupling an MZM to the dot of the DRLM [see Fig.~\ref{fig:phase_diagrams} that is distinct from Fig.~\ref{figure:model}(b)].
With Coulomb gas renormalization group (RG) equations and g-theorem~\cite{AffleckLudwigPRL91,AffleckLudwigAndreasPRB93,FriedanKonechnyPRL04}, we further prove that the relaxation of symmetry requirement is a direct consequence of the
topological degeneracy, with which the left-right asymmetry becomes irrelevant near the strong-tunneling fixed point.
Our work, as far as we know, is pioneering in studying the interplay between
topological degeneracy and boundary QPTs where symmetries are otherwise strictly requested.
In addition, the broadened ballistic-transmission phase space (Fig.~\ref{fig:phase_diagrams}) greatly reduces the experimental difficulty to observe the BKT QPT of a DRLM.
Remarkably, the phase diagram featuring MZMs, which greatly contrasts that of a regular DRLM, provides a clear and unequivocal indicator for detecting the presence of MZMs. 
In real experiments, to observe our predicted phenomenon (i.e., phase diagrams in Figs.~\ref{figure:model} and \ref{fig:phase_diagrams}), the temperature and applied voltage-bias are required to be smaller than the Majorana-dot coupling, and effective lead-dot hybridizations.

\textbf{\emph{Model}}---
The system we consider [with the Hamiltonian $H = H_\text{leads} + H_\text{impurity} + H_\text{T}$, see Fig.~\ref{figure:model}(a)] contains two dissipative leads ($H_\text{leads}$) that couple to ($H_\text{T}$) a Majorana-contained impurity part $H_\text{impurity} = \epsilon_d d^\dagger d + i t_\text{M} (d + d^\dagger) \gamma_1/\sqrt{2} $,
where $d$ indicates the dot that couples to one MZM $\gamma_1$.
The other MZM $\gamma_2$ [not shown in Fig.~\ref{figure:model}(a)] decouples from $\gamma_1$, to enforce the exact topological degeneracy.
For later convenience, we define an auxiliary fermionic operator $d_\text{MZM} \equiv (\gamma_1 - i\gamma_2)/\sqrt{2}$ with MZMs.
The lead Hamiltonian $H_\text{leads}$ contains two free fermionic leads, and bosonic environmental modes.
With those modes included, the lead-impurity tunneling Hamiltonian $H_\text{T} = V_L \psi^\dagger_L d e^{i\varphi_L} + V_R \psi^\dagger_R d e^{i\varphi_R} + h.c.$ contains the dynamical phase $\varphi_\alpha$ that is conjugate to the charge number $N_\alpha = \int dx \psi^\dagger_\alpha(x) \psi_\alpha (x)$, in the corresponding lead $\alpha = L, R$.
After integrating out the environmental modes, $\varphi_L = - \varphi_R = \varphi$, with the long-time dynamical feature $\langle \varphi (\tau) \varphi(0) \rangle \sim 2r \ln \tau$~\cite{Ingold1992SUS}, where $r = \mathcal{R} e^2/h$ is the ratio between the Ohmic impedance $\mathcal{R}$ of the dissipative leads, and the resistance quanta $h/e^2$.
The required dissipation has been realized in real experiments~\cite{PierreNatPhys11,Mebrahtu12,Mebrahtu13,JezouinPierre13}.
Finally, two leads couple to the dot, with hybridization parameters $V_L$ and $V_R$.

Dissipative modes introduce effective many-body interaction to the system~\cite{NazarovJETP89,SafiSaleurPRL04,FlorensPRB07}.
We then bosonize lead fermions after extending semi-infinite leads into chiral full-1D ones~\cite{KaneFisherPRB92,GiamarchiBook} $\psi_\alpha (x) = F_\alpha \exp [i\phi_\alpha (x)]/\sqrt{2\pi a}$,
where $a$ is the short-distance cutoff, and $F_\alpha$ is the Klein factor that enforces the fermionic anti-commutator. Finally, the bosonic phase $\phi_\alpha$ satisfies the commutation relation $[\phi_\alpha (x), \phi_{\alpha'} (x')] = i\pi \Theta (x - x') \delta_{\alpha, \alpha'}$.
These fields are related to the charge numbers in two leads $N_{L,R} = e [\phi_{L,R} (\infty) - \phi_{L,R} (-\infty) ]/(2\pi)$.
To proceed, we define the common and difference fields with the rotation $\phi_{c,f} = (\phi_L \pm \phi_R)/\sqrt{2}$. The post-rotation fields $\phi_c$ and $\phi_f$ reflect the total charge number $N_L + N_R$, and the number difference $N_L - N_R$, respectively.
With bosonization and the rotation, the dissipative phase can be combined with the difference field $\phi_f' \equiv (\phi_f + \varphi/\sqrt{2})/\sqrt{1 + r}$. The addressed dissipative phase $\varphi' \equiv (\sqrt{r} \phi_f - \varphi/\sqrt{2r})/\sqrt{1 + r}$ decouples, reducing the system degrees of freedom.

\textbf{\emph{BKT QPT of the DRLM}} --- When the MZM decouples ($t_\text{M} = 0$), the system reduces to the regular DRLM, where a boundary BKT QPT is predicted~\cite{DongHuaixiuPRB14} when $2 < r <3$ and $V_L = V_R$~\cite{Footnote} (see an alternative QPT interpretation in Ref.~\cite{ChunghouPRL09}).
When $V_L = V_R$ is larger than a critical value~\cite{SupMat}, the system flows to the ballistic-transmission phase with conductance $e^2/h$ [Line B in the phase diagram of Fig.~\ref{figure:model}(b)].
Notice that this conductance is different from that of Ref.~\cite{DongPRB11,VernekPRB14}, due to the dissipative and interacting effect~\cite{SafiSchulzPRB95,MaslovStonePRB95,PonomarenkoPRB95,Arisato96,ChamonEduardoPRB97,Mebrahtu12,HuKaneX16}.
By contrast, with couplings smaller than the critical value, the system instead flows to the isolated-impurity phase [A of Fig.~\ref{figure:model}(b)] with zero conductance.
The transition (between phases A and B) is of BKT-type, due to the absence of QCP.
Indeed, it shares equivalent scaling equations as that of an XY-model, the pioneering model of topological transition~\cite{Berezinskii71,Kosterlitz74,KosterlitzThouless18}.
More specifically, when $r > 2$, the lead-dot coupling is originally RG irrelevant [i.e., with scaling dimension larger than one~\cite{AltlandSimonsBook}, see Eq.~\eqref{eq:high-t-flow}].
The couplings then, initially, decrease during the RG flow.
If $V_L = V_R$ are initially large enough, the dot charge number gradually becomes fixed at half-filling~\cite{KaneFisherPRB92,DongHuaixiuPRB14}.
The common field, reflecting the charge number fluctuation, then loses its scaling dimension after the RG, leading to a relevant lead-dot coupling if $r < 3$.
Briefly, the change of scaling dimension (upon the RG flow) leads to the BKT QPT: the system can enter the ballistic-transmission phase, only if $V_L$ and $V_R$ become relevant before they vanish; otherwise the flow terminates in the isolated-impurity phase.

Importantly, this BKT QPT requires a strict left-right symmetry $V_L = V_R$.
Indeed, if $V_L \neq V_R$, the system flows to other two zero-conductance phases [C and D of Fig.~\ref{figure:model}(b)].
The transition $\text{C}\leftrightarrow \text{B} \leftrightarrow \text{D}$, with state B as the QCP, is not of BKT type.
As our main result, we show that with a finite Majorana-dot coupling $t_\text{M}$, the strict left-right symmetry to observe the BKT QPT is relaxed.
Indeed, as shown in Fig.~\ref{fig:phase_diagrams}, the ballistic-transmission phase (area B) has extended from a 1D line [Fig.\ref{figure:model}(b)] to a 2D space.
The specific phase diagram changes under the different values of $t_\text{M}$: when $t_\text{M}$ decreases 
[i.e. increasing cutoff $l_C$, from Fig.~\ref{fig:phase_diagrams}(a) to Fig.~\ref{fig:phase_diagrams}(b)], 
the ballistic-transmission phase requires a smaller lead-dot tunnelings, but has a worse tolerance against left-right asymmetry.

\textbf{\emph{Coulomb gas RG}} ---
In this work, we obtain the phase diagram of Fig.~\ref{fig:phase_diagrams} (a,b)
with the method of Coulomb gas RG~\cite{AndersonPRB70,KaneFisherPRB92}.
Briefly, within the Coulomb gas representation, one works with the partition function
\begin{equation}
\begin{aligned}
\mathcal{Z} &= \sum_n\int_0^{1/T} \!d\tau_1\! \cdots \int_0^{\tau_{i}-\tau_c} \!d\tau_{i+1}  \!\cdots \! \!\int_0^{\tau_{n - 1}-\tau_c} \! d\tau_{n} \\
& \big\langle H_{\text{T}}(\tau_{1}) \cdots H_{\text{T}}(\tau_{i+1})  \cdots H_{\text{T}}(\tau_{n}) \big\rangle_{H_\text{leads} + H_\text{impurity}} ,
\end{aligned}
\label{eq:partition_function}
\end{equation}
where $n$ refers to the number of lead-impurity tunneling events during imaginary time $1/T$, and $\tau_c$ is the cutoff: for any $i \in [1, n-1]$, $\tau_i - \tau_{i+1}  > \tau_c$ is enforced.
Expansions of Eq.~\eqref{eq:partition_function} are carried out within the interaction picture~\cite{FlensbergBook}, over lead-impurity tunnelings $H_\text{T}$.
One can perform RG calculations~\cite{KaneFisherPRB92} as follows.
During each RG step, one starts with the cuttoff $\tau_c$, and increases it to $\tau_c + d\tau_c$ (indicating the temperature decreasing). The pairs of tunneling events at $\{\tau_i, \tau_{i+1}\}$ with condition $\tau_c < |\tau_i - \tau_{i+1} | < \tau_c + d\tau_c$ should be integrated out. RG flow equations are then obtained, after rescaling the cutoff, and enforcing the invariance of $\mathcal{Z}$.
In Eq.~\eqref{eq:partition_function}, the correlator of $H_\text{T}$ equals the product of the free-lead correlation (with the environmental modes included) and the impurity correlation.
The free-lead contribution can be straightforwardly obtained~\cite{KaneFisherPRB92,DongHuaixiuPRB14}.

The impurity correlator can be obtained by solving the Hamiltonian $H_\text{impurity}$, leading to four impurity eigenstates: two even-parity ones $|\psi_{1,3} \rangle = \frac{1}{\sqrt{2}} (\mp i|1,1\rangle + | 0,0\rangle)$, and two odd-parity ones $|\psi_{2,4} \rangle = \frac{1}{\sqrt{2}} (\mp i|1,0\rangle + | 0,1\rangle)$,
where $|n,m\rangle$ indicates the state with $n$ and $m$ particle (either 0 or 1) in the dot and the auxiliary dot $d_\text{MZM}$, respectively.
Assuming $t_\text{M} >0$, states $|\psi_1\rangle$ and $|\psi_2\rangle$ become the degenerate ground states: they have the energy $-t_\text{M}$, lower than that ($t_\text{M}$) of $|\psi_3\rangle$ and $|\psi_4\rangle$.

Without lead-dot tunneling, the impurity part stays in two possible ground states $|\psi_1\rangle$ and $|\psi_2\rangle$.
This double-degeneracy originates from the topological degeneracy.
After one lead-dot tunneling, they become $d |\psi_1\rangle \! = \! - \!\frac{i}{2} (|\psi_2\rangle \!+\! |\psi_4\rangle), d^\dagger |\psi_1\rangle \!=\! \frac{i}{2} (|\psi_2\rangle \!-\! |\psi_4\rangle),  d |\psi_2\rangle  \!=\! -\!\frac{i}{2} (|\psi_1\rangle \!+\! |\psi_3\rangle),$ and $d^\dagger |\psi_2\rangle \!=\! \frac{i}{2} (|\psi_1\rangle \!-\! |\psi_3\rangle)$, resulting in (non-eigen) final states with impurity parities different from the initial ones.
The impurity state before the next lead-impurity tunneling depends on the evolution time~\cite{SupMat}.
For the low-temperature limit $T \ll t_\text{M}$, tunnelings are rare, the impurity state evolves back to the ground states $|\psi_1\rangle$ and $|\psi_2\rangle$ before the next tunneling.
By contrast, for the high-temperature situation $T \gg t_\text{M}$, lead-dot tunnelings are frequent, and the impurity state stays in the non-eigenstate before the next lead-impurity tunneling.
We thus have to treat the low-temperature and high-temperature situations separately.

\textbf{\emph{The high-temperature RG flow}} --- We start by visiting the high-temperature regime $T \gg t_\text{M}$, where all four impurity states are energetically allowed.
Of this case, the impurity-lead tunneling histories
with alternating creation and annihilation operators dominate the partition function~\cite{SupMat}.
The effect of the MZM-dot coupling is then negligible, leading to RG equations that approximately equal that of a normal DRLM (following Ref.~\cite{DongHuaixiuPRB14}, or reducing Luttinger liquid spinful results of Ref.~\cite{SchillerIngersentRG97} to the spinless case),
\begin{equation}
\begin{aligned}
& d V_{L,R}/\! dl \! = \! V_{L,R} \left[ 1 \! - \! \frac{1+r}{4} (1\! +\! K_\text{charge} \!\pm \! 2 K_\text{asy}) \right],\\
& d K_\text{charge}/ \! dl \! = \! - 4 \tau_c^2 [K_\text{charge} (V_L^2 \! + \! V_R^2) \!+\! K_\text{asy}  (V_L^2 \!-\! V_R^2) ], \\
& d K_\text{asy}/ \! dl \!=\! -2 \tau_c^2 [K_\text{asy} (V_L^2 \!+\! V_R^2) \!+\! (V_L^2 \!-\! V_R^2) ],
\end{aligned}
\label{eq:high-t-flow}
\end{equation}
where parameter $l$ indicates the RG flow progress~\cite{SupMat}: it equals zero at the beginning of RG, and increases when temperature $T$ decreases $l\sim \ln(T_0/T)$, with $T_0$ the initial temperature (a device-dependent parameter that can only be characterized experimentally).
$K_\text{charge}$, with its initial value $1/(1+r)$, refers to the scaling dimension of the common field $\phi_c$.
$K_\text{asy}$, which equals zero initially, refers to the scaling dimension from the left-right asymmetry~\cite{KaneFisherPRB92,DongHuaixiuPRB14}.
These two parameters are introduced to model the change of bosonic-field correlation functions, during the RG flow.
Briefly, these two parameters are introduced to model two influences of the RG flow. Firstly, during the RG flow, common field correlation, which equals $\langle \exp [i\phi_c(t)] \exp [-i\phi_c(0)] \rangle_\text{initial} \propto 1/t$ initially, becomes modified into the effective correlation $\langle \exp [i\phi_c(t)] \exp [-i\phi_c(0)] \rangle_\text{RG-effective} \propto 1/t^{K_\text{charge}}$.
As the second effect, if $V_L \neq V_R$, the common ($\phi_c$) and difference ($\phi_f$) fields, which were initially uncorrelated, obtain the effective correlation $\langle \exp [i\phi_c(t)] \exp [-i\phi_f(0)] \rangle_\text{RG-effective} \propto 1/t^{K_\text{asy}}$, leading to, crucially, different correlations of bosonic fields ($\phi_L$ and $\phi_R$).
Correlations above, evaluated during the RG flow, on the other hand define these two quantities $K_\text{charge}$ and $K_\text{asy}$.
Crucially, these two parameters are conventionally involved in studying interacting systems (e.g., Refs.~\cite{KaneFisherPRB92,SchillerIngersentRG97}).
For the symmetric case $V_L = V_R$, $K_\text{asy}$ is fixed at zero [red dashed line of Fig.~\ref{fig:phase_diagrams}(c)].
The scaling dimension of lead-dot tunnelings becomes smaller than one (thus RG relevant) when $K_\text{charge}$ has become smaller than $K_C = 4/(1+r) - 1$.
For asymmetric situations, by contrast, $|K_\text{asy}|$ increases during the RG flow [blue dashed line of Fig.~\ref{fig:phase_diagrams}(c)], after which the larger lead-dot coupling becomes more relevant than the weaker one, leading to an even more significant asymmetry~\cite{SupMat}.
The high-temperature flow terminates when $T\sim t_\text{M}$.

\textbf{\emph{Low-temperature RG, and the phase diagram}} --- For the low-temperature case $T \ll t_\text{M}$, only 
ground states are allowed as real states in the impurity, and after one lead-dot tunneling, the impurity system prefers to evolve to one of the ground states ($|\psi_1\rangle$ or $|\psi_2\rangle$), before the next lead-dot tunneling event.
Neighbouring dot operators can be thus the same: a consequence of the presence of topological degeneracy.
Low-temperature RG equations of an MZM-involved system become~\cite{SupMat}
\begin{equation}
\begin{aligned}
    & d V_{L,R}/dl \!=\! V_{L,R} \left[ 1 \! - \! \frac{1+r}{4} (1\! +\! K_\text{charge} \!\pm \! 2 K_\text{asy}) \right],\\
    &d K_\text{charge}/dl = 0, \ \ d K_\text{asy}/dl = -2\tau_c^2 (V_L^2 + V_R^2) K_\text{asy},
\end{aligned}
\label{eq:low-t-flows}
\end{equation}
with initial values of $K_\text{charge}$ and $K_\text{asy}$, importantly, inherited from the RG-flow of the previous stage.
The low-temperature flow Eq.~\eqref{eq:low-t-flows} contains two major Majorana-induced modifications: (i) $K_\text{charge}$, which initially decreases in the high-temperature situation [solid lines of Fig.~\ref{fig:phase_diagrams}(c)], now stops flowing [solid lines of Fig.~\ref{fig:phase_diagrams}(d)], and, more importantly, (ii) $K_\text{asy}$, which increases in the asymmetric case of the high-temperature limit [dashed red line of Fig.~\ref{fig:phase_diagrams}(c)], now instead begins to decrease to zero [dashed lines of Fig.~\ref{fig:phase_diagrams}(d)].
Modification (ii), remarkably, is the key element that leads to the irrelevance of a weak asymmetry. Indeed, when $K_\text{asy}$ has vanished, $d V_{L,R}/dl \!=\! V_{L,R} \left[ 1 \! - \! (1+r) (1\! +\! K_\text{charge} )/4 \right]$, where $V_L$ and $V_R$ simultaneously increase to perfect or decrease to zero, disregarding a weak asymmetry.
Noticeably, the irrelevance of asymmetry also removes phases C and D in Fig.~\ref{figure:model}(c), where only one lead-impurity coupling (the smaller one of $V_L$ or $V_R$) flows to zero.
As the consequence of (i), the final phase is determined by the value of $K_\text{charge}$, larger or smaller than $K_C$, at the end of high-temperature flow [as shown in Fig.~\ref{fig:phase_diagrams}(d)].

\begin{figure}
  \includegraphics[width=\linewidth]{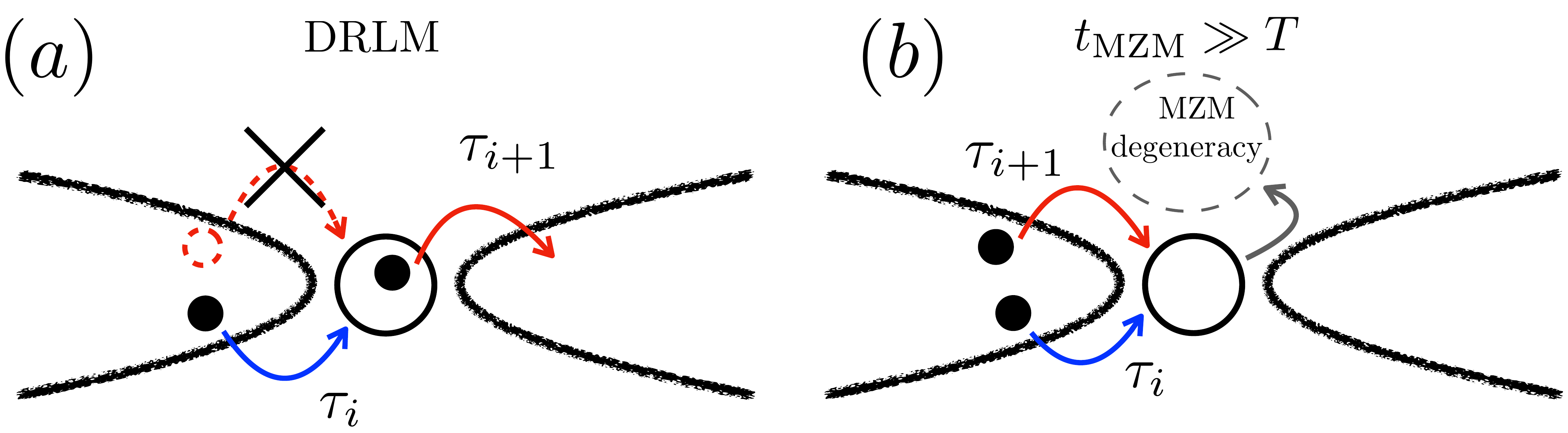}
  \caption{Lead-impurity tunneling events. (a) In a regular DRLM, the dot state alternates (between two degenerate dot states) in time. (b) In great contrast, with tunneling into the MZM degeneracy (the grey arrow), two neighbouring dot-lead tunneling events (shown by the blue and red arrows) can both contain a creation operator in the dot.}
  \label{fig:field_lattices}
\end{figure}

One can physically understand these two modifications with Fig.~\ref{fig:field_lattices}.
Briefly, in DRLM [with corresponding RG equations~\eqref{eq:high-t-flow}] the dot state alternates in time [Fig.~\ref{fig:field_lattices}(a)], where $N_L + N_R$ has only two possible (integer) options. Field $\phi_c$, reflecting fluctuations of $N_L + N_R$, then gradually loses its scaling dimension [reflected by the decreasing $K_\text{charge}$ of Eq.~\eqref{eq:high-t-flow}] when $N_L + N_R$ is fixed by a half-filling dot~\cite{KaneFisherPRB92}.
In great contrast, with Majorana involved, the parity of the impurity part, including the dot and a coupled MZM, alternates in time [Fig.~\ref{fig:field_lattices}(b)].
The corresponding common field $\phi_c$ now becomes a free 1D field.
Its scaling dimension $K_\text{charge}$ thus remains invariant during RG flow of Eq.~\eqref{eq:low-t-flows}.
We emphasize, importantly, the necessity of topological degeneracy in the picture above.
Instead, if the DRLM dot couples to a zero-energy ABS, 
the case can reduce to the DRLM with a detuned dot~\cite{SupMat}: a system with only the isolated-impurity phase~\cite{Mebrahtu12,Mebrahtu13,DongHuaixiuPRB14}.
This effective detuning is proportional to the difference in amplitudes of normal and Andreev tunnelings between ABS and the dot.
Crucially, assuming the effective detuning $\varepsilon$ (not the inter-MZM coupling that can be reduced by increasing the nanowire size), it obeys~\cite{SupMat}
\begin{equation}
    \frac{d}{dl}\varepsilon = \varepsilon + \frac{\tau_c (V_L^2 + V_R^2)}{4} (e^{- \varepsilon\tau_c} - e^{ \varepsilon \tau_c}), 
    \label{eq:abs_energy}
\end{equation}
and grows during the RG flow.
Consequently, however small initially, $\varepsilon$ becomes dominant, driving the system to the zero-conductance phase at energies much smaller than $\varepsilon$.
Specially, $\varepsilon$ equals zero initially, for an accidentally fine-tuned ABS. It however will not provide an extended ballistic-transmission phase, due to inevitable cross-talks between tuning gates~\cite{SupMat}.

Modifications above also explain features of Fig.~\ref{fig:phase_diagrams}(a,~b).
More specifically, with a larger $t_\text{M}$ of Fig.~\ref{fig:phase_diagrams}(a), (i) for the symmetric case, a larger value of $V_L = V_R$ is required to enter the ballistic-transmission phase, and (ii) for the asymmetric case, the ballistic-transmission phase tolerates a larger value of $V_L - V_R$.
Feature (i) is a consequence of an earlier termination, due to a larger $t_\text{M}$, of the high-temperature flow.
Feature (ii) is related to the non-monotonous curve of $K_\text{charge}$ [blue solid line of Fig.~\ref{fig:phase_diagrams}(c)].

\begin{figure}
  \includegraphics[width=\linewidth]{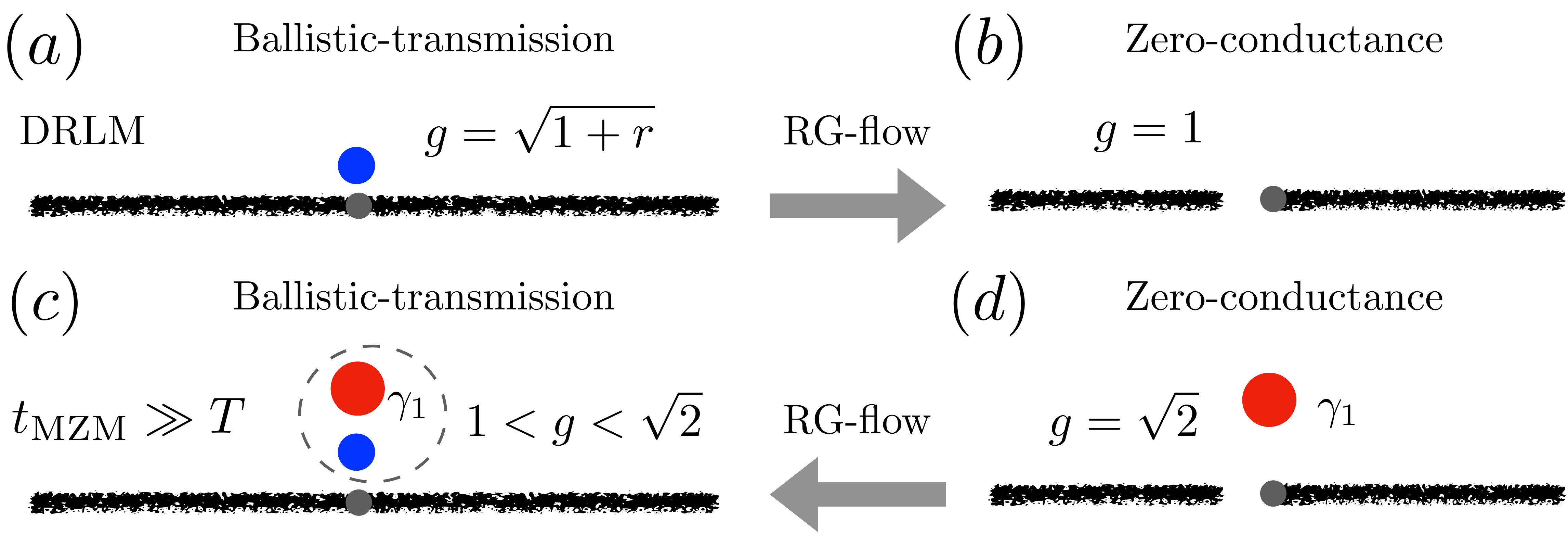}
  \caption{The effect of left-right asymmetry (assuming $V_L < V_R$) in a regular [(a) and (b)] and Majorana-involved DRLM [(c) and (d)].
  The red and blue dots refer to the MZM and the residue degeneracy ($g = \sqrt{1 + r}$) of the ballistic-transmission phase, respectively.
  The grey dot instead indicates the hybridized dot degree of freedom.
RG flow prefers the fixed point with a smaller degeneracy.
  }
  \label{fig:asymmetry_effect}
\end{figure}

\textbf{\emph{Discussions}} --- 
In this paper, we show that the MZM-introduced topological degeneracy can relax the strict symmetry requirement $V_L = V_R$ for a BKT QPT in a DRLM.
This result can be understood with the g-theorem~\cite{AffleckLudwigPRL91,AffleckLudwigAndreasPRB93}, by visiting the effect of a weak asymmetry in the strong tunneling regime. Briefly, the g-theorem claims that in a boundary QPT, the phase with a smaller degeneracy is more stable.
Indeed, in a regular asymmetry-present DRLM, the fixed point where the dot is fully hybridized by the stronger lead has a smaller degeneracy [degeneracy $g = 1$, in Fig.~\ref{fig:asymmetry_effect}(b)] than that of the ballistic-transmission fixed point [$g = \sqrt{1 + r}$~\cite{HuaixiuSergeHarold}, Fig.~\ref{fig:asymmetry_effect}(a)].
The zero-conductance phase is thus more stable, in a regular DRLM.
With Majorana provided,
the fixed point where dot is hybridized by the stronger lead has the degeneracy $\sqrt{2}$ [Fig.~\ref{fig:asymmetry_effect}(d)], the degeneracy of a single MZM: larger than that [Fig.~\ref{fig:asymmetry_effect}(b)] of the MZM absent situation.
For the ballistic transmission case, the Majorana couples to the residued dot-degeneracy (i.e., $\sqrt{1 + r}$).
The degeneracy of the impurity part, obtained after the MZM-residue fusion~\cite{AppliedCFT88,CFTBook} (see Ref.~\cite{AffleckLudwigJonesPRB95}, a pioneering example), is generally hard to obtain.
Indeed, an arbitrary $r$ does not necessarily correspond to a discrete Virasoro algebra with central charge $c < 1$~\cite{AppliedCFT88}.
We can however evaluate the possible range of degeneracy, with fusion between primary fields of the Ising model ($c = 1/2$). Indeed, if $r = 1$, we have the fusion $[\sqrt{1 + r}] \times [\sqrt{2}] = [\sqrt{2}] \times [\sqrt{2}] = [1] + [2]$, with numbers indicating the degeneracy. For $r = 3$, instead $[\sqrt{1 + r}] \times [\sqrt{2}] = [2] \times [\sqrt{2}] = [\sqrt{2}] + [2\sqrt{2}]$.
Taking the smaller degeneracy of both cases,
the Majorana-involved ballistic-transmission phase thus has a smaller degeneracy [$1 < g < \sqrt{2}$, Fig.~\ref{fig:asymmetry_effect}(c)] than that ($\sqrt{2}$) of the zero-conductance one, indicating a stabilized ballistic-transmission phase against a weak asymmetry.
Importantly, similar g-theorem analysis applies to situations with a finite $\epsilon_d$: the BKT QPT thus tolerates also a small dot detuning~\cite{SupMat}.

Before closure,
we emphasize that the predicted modification of the BKT QPT phase diagram can further confirm plausible Majorana signature (if any), through the detection of the non-trivial MZM degeneracy.
Indeed, with a zero-energy ABS (which notoriously reduces the reliability of Majorana signals), the impurity part, including the resonant level and the ABS, is gapped in energy [i.e., $\varepsilon$ of Eq.~\eqref{eq:abs_energy}], resulting into a suppressed current at low energies.
A detailed calculation and corresponding g-theorem analysis are provided in the Supplementary Information~\cite{SupMat}.
Remarkably, conclusions above even remain applicable to situations where the hybrid nanowire has multiple conductance channels.
Encouraged by persistent progress of Majorana hunting in nano-deveices~\cite{mourik2012signatures,deng2016majorana,Albrecht2016Nature,fornieri2019evidence,ren2019topological,vaitiekenas2021zero,song2022large,WangPRL22,aghaee2022inas,Dvir2023Nature}, especially for the cases with dissipative environments~\cite{liu2013proposed,DonghaoPRL22,LargeDissipation22,WeakDissipation22,zhang2022situ},
we anticipate the experimental capability to introduce enough dissipation in potential Majorana-hosted candidates.
Prospectively, other QPT-hosted systems (either one-dimensional ones, or systems with multiple boundary QPTs) can potentially help the identification of Majorana non-localities~\cite{SemenoffSodanoX06,Fu2010PRL,Beri&Cooper-TKE,Altland&Egger,BeriPRL2013,AltlandPRL14,HaoNatCom22}, another crucial elements in topological quantum device.

\emph{Acknowledgements} ---
The authors thank Hao Zhang for valuable discussions. This work was supported by the Innovation Program for Quantum Science and Technology (Grant No.~2021ZD0302400), the National Natural Science Foundation of China (Grants No.~11974198, No.~12374158, and No.~12074039).

\widetext
\clearpage

\renewcommand{\bibnumfmt}[1]{[S#1]}
\renewcommand{\citenumfont}[1]{S#1}
\global\long\def\theequation{S\arabic{equation}}
\global\long\def\thefigure{S\arabic{figure}}
\setcounter{equation}{0}
\setcounter{figure}{0}

\begin{center}
\textbf{\large Supplementary Information for ``Influence of topological degeneracy on the boundary Berezinskii-Kosterlitz-Thouless quantum phase transition of a dissipative resonant level''\\
	\vspace{5pt}}
\vspace{15pt}
Gu Zhang, Zhan Cao, and Dong E. Liu\\
(Dated: April 26th, 2024)
\end{center}
\vspace{15pt}

In this Supplementary Information, we provide details on: (i) RG equations and descriptions of BKT QPT in a regular DRLM, (ii) correlation function of the impurity part, (iii) derivations of RG equations in the low-temperature regime, (iv) RG flow equations in the presence of a finite detuning of the dot ($\epsilon_d$), (v) the situation that the DRLM dot couples to a topologically trivial Andreev bound state, and (vi) a proposed protocol, assisted by the BKT transition, to further strength the credibility of a plausible Majorana signal.

\section{BKT transition in DRLM}

\begin{figure}
  \includegraphics[width = 0.6 \linewidth]{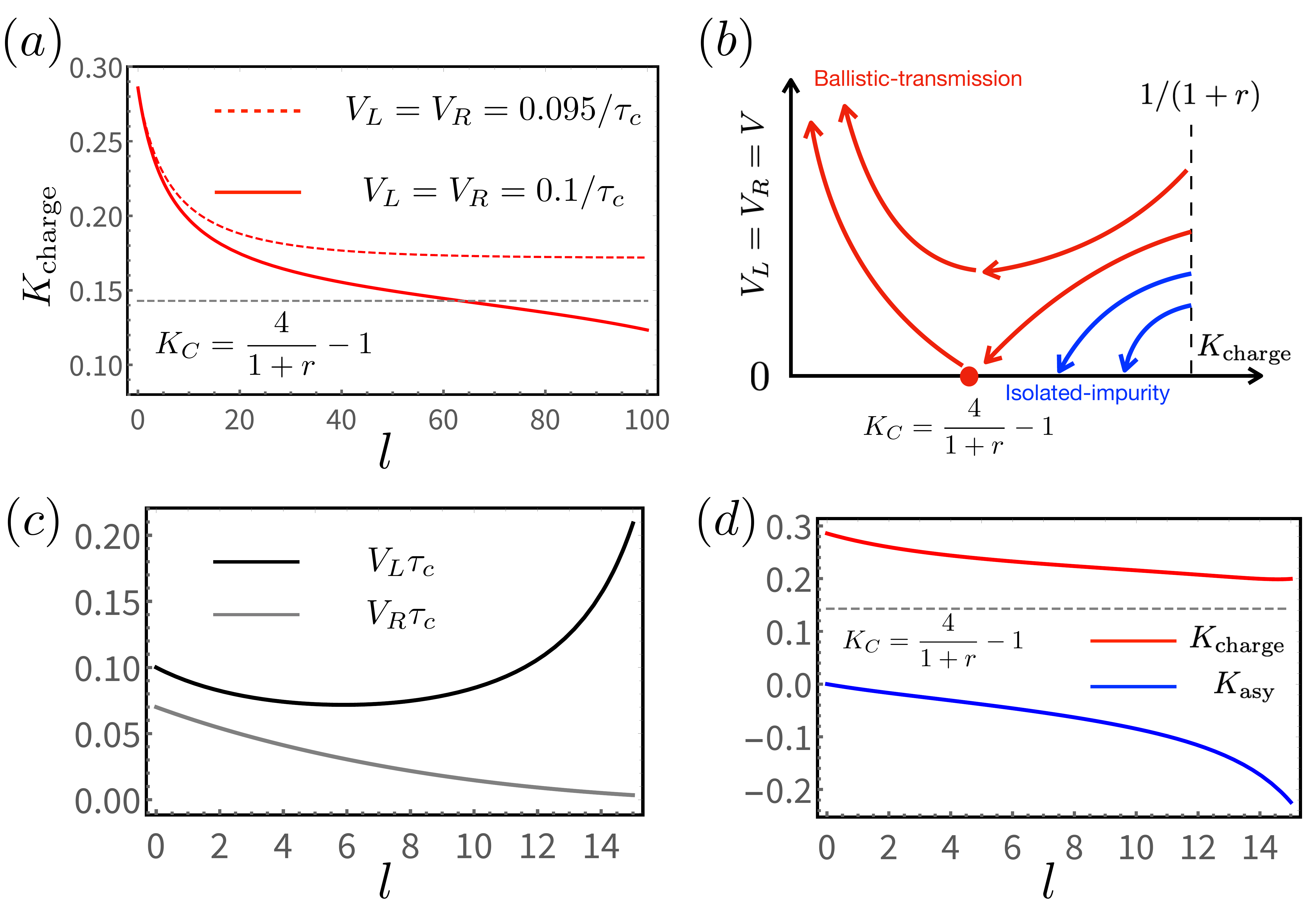}
  \caption{RG flows of a DRLM, when $r=2.5$.
  (a) The flow of $K_\text{charge}$, when $V_L = V_R = 0.095 \tau_c$ (the dashed line) and $V_L = V_R = 0.1 \tau_c$ (the solid line), respectively. Of the latter case, $K_\text{charge}$ flows beneath the critical value $K_C = -1 + 4/(1 + r)$, leading to the ballistic-transmission phase. By contrast, of the former case, $K_\text{charge}$ saturates to a value larger than $K_C$, leading to the isolated-impurity phase. (b) The corresponding RG flow diagram~\cite{SDongHuaixiuPRB14}.
  It greatly mimics that of a Kondo model, where a BKT QPT is predicted between the fully screened and the isolated-impurity phases~\cite{SHewsonBook}.
  (c) The flow of $V_L$ and $V_R$ if they start with different initial values. $V_L$ becomes relevant (i.e., increases with an increasing $l$) after the RG flow. The weaker one $V_R$ however remains irrelevant. (d) The flow of $K_\text{charge}$ and $K_\text{asy}$, with the same parameters in (c). $V_L$ becomes relevant when $K_\text{charge} > K_C$, due to a finite $K_\text{asy}$.
  }
  \label{fig:drlm}
\end{figure}

In this section, we provide more details to present the existence of a BKT QPT in DRLM.
We start with the DRLM RG flow equation, Eq.~(5) of the main text,
\begin{equation}
\begin{aligned}
& \frac{d V_{L,R}}{dl} = V_{L,R} \left[ 1 \! - \! \frac{1+r}{4} (1\! +\! K_\text{charge} \!\pm \! 2 K_\text{asy}) \right],\\
& \frac{d K_\text{charge}}{dl} \! = \! -4 \tau_c^2 [K_\text{charge} (V_L^2 \! + \! V_R^2) \!+\! K_\text{asy} (V_L^2 \!-\! V_R^2) ], \\
& \frac{d K_\text{asy}}{dl} \!=\! -2 \tau_c^2 [K_\text{asy} (V_L^2 \!+\! V_R^2) \!+\! (V_L^2 \!-\! V_R^2) ].
\end{aligned}
\label{eq:drlm-flow}
\end{equation}
Without the Majorana, the flow of Eq.~\eqref{eq:drlm-flow} continues until arriving at the cutoff: either temperature of bias.
For the symmetric case $V_L = V_R$, we obtain the flow of $K_\text{charge}$ in Fig.~\ref{fig:drlm}(a).
Here the dashed and solid red lines have $V_L = V_R = 0.095/\tau_c$ and $V_L = V_R = 0.1/\tau_c$, respectively. The dashed grey line indicates the critical value [$K_C = 4/(1 + r) - 1$] of $K_\text{charge}$: lead-dot tunnelings become relevant when $K_\text{charge} < K_C$, and irrelevant given the opposite situation, i.e., $K_\text{charge} > K_C$.
If $K_\text{charge} > K_C$ during the entire RG flow [the red dashed line of Fig.~\ref{fig:drlm}(a)], couplings of the dot to both leads are always irrelevant, and the system flows to the isolated-impurity phase with zero equilibrium conductance.
By contrast, if $K_\text{charge} < K_C$ after the RG flow [the red solid line of Fig.~\ref{fig:drlm}(a)], system flows instead to the ballistic-transmission phase with perfect conductance $e^2/h$.
Analysis above leads to the RG flow diagram of Ref.~\cite{SDongHuaixiuPRB14} (one can see Ref.~\cite{SChunghouPRL09} for an alternative interpretation of this BKT QPT), shown in Fig.~\ref{fig:drlm}(b).

Now we move to discuss the influence of asymmetry. Figs.~\ref{fig:drlm}(c) and \ref{fig:drlm}(d) present the flow of dot-lead couplings and scaling parameters ($K_\text{charge}$ and $K_\text{asy}$), respectively.
As the stronger lead-dot tunneling, $V_L$ becomes RG relevant (i.e., increases with an increasing $l$) after the flow.
The weaker one however remains irrelevant.
In (d), we further see that $V_L$ becomes relevant when $K_\text{charge} > K_C$, due to a finite $K_\text{asy}$.

\section{Correlation function of the impurity part}

In this section we provide detailed discussions on the eigenstates of the impurity part, and the evolution of impurity states after lead-impurity tunnelings.
As a reminder, in this work we consider the impurity Hamiltonian
\begin{equation}
    H_\text{impurity} = \epsilon_d d^\dagger d + i t_\text{M} (d + d^\dagger) \gamma_1/\sqrt{2},
\end{equation}
with $d$ and $d^\dagger$ the dot operators, and $\gamma_1$ representing for the Majorana.
As has been shown in the main text, in this work we take the Coulomb gas RG method, and work in the interaction picture that contains expansions over lead-impurity tunnelings.
The corresponding correlation functions then contain the free-lead part and the isolated impurity part. The former one is straightforward within the Luttinger liquid formalism~\cite{SKaneFisherPRB92}, while the latter one equals
\begin{equation}
\begin{aligned}
 \big\langle \Pi_{i = 1}^n D(\varepsilon_i,\tau_i) \big\rangle =  2 \Pi_{i = 1}^{n-1} g_\text{dot} (\tau_i - \tau_{i+1}, \varepsilon_i, \varepsilon_{i+1}),
\end{aligned}
\label{eq:dot_correlation}
\end{equation}
where we have, for convenience, defined another dot operator $D(\varepsilon_i,\tau_i)$: it equals $d^\dagger (\tau_i)$ and $d (\tau_i)$ if $\varepsilon = \pm 1$, respectively.
The green's function $g_\text{dot} (\tau,\varepsilon, \varepsilon') = [\exp(t_\text{M} \tau)  - \varepsilon \varepsilon' \exp(-t_\text{M} \tau)]/2$: it equals $\sinh (t_\text{M} \tau)$ if dot operators of two involved neighbouring kinks are both creation or annihilation operators (i.e., $\varepsilon_i \varepsilon_{i+1} = 1$); otherwise ($\varepsilon_i \varepsilon_{i+1} = -1$) it equals $\cosh (t_\text{M} \tau)$.
One can understand RG flows under different temperatures by visiting the corresponding features of Eq.~\eqref{eq:dot_correlation}.
Briefly, if temperature is large,
$\tau < 1/T \ll 1/t_\text{M}$, $\sinh ( t_\text{M} \tau ) \ll \cosh ( t_\text{M} \tau )$, and lead-tunneling histories 
with alternating dot states (i.e., with $\varepsilon_i \varepsilon_{i+1} = -1$) are preferred. The correlation function of the impurity part then greatly mimics that of a regular DRLM.
By contrast, for a small temperature, $\tau \gg 1/t_\text{M}$ in general, $\sinh (t_\text{M} \tau ) \approx \cosh(t_\text{M} \tau )$, and now the parity of the entire impurity part alternates in time, leading to RG flows that are distinct from that of a regular DRLM.

One can similarly understand the difference between high-temperature and low-temperature situations, by looking into the time-dependent evolution of the impurity states.
Without loss of generality, we assume the state  $|\psi_{2} \rangle = \frac{1}{\sqrt{2}} (- i|1,0\rangle + | 0,1\rangle)$ of the impurity at a moment.
After receiving one electron via a lead-impurity tunneling, impurity state becomes $|1,1\rangle$.
As a non-eigen state, $|1,1\rangle$ evolves in time, becoming $\exp(- H_\text{impurity} \tau) |1,1\rangle= \cosh(t_\text{M} \tau) | 1, 1 \rangle + i \sinh(t_\text{M} \tau)| 0, 0 \rangle$, after $\tau$ of the tunneling.
For the high-temperature limit, tunnelings are frequent, $t_\text{M} \tau \ll 1$ is a small quantity. The impurity state $\cosh(t_\text{M} \tau) | 1, 1 \rangle + i \sinh(t_\text{M} \tau)| 0, 0 \rangle \approx |1,1\rangle$ approximately equals that right after the lead-impurity tunneling.
By contrast, for the low-temperature limit, tunnelings are rare, $t_\text{M} \tau \gg 1$ is normally large. Of this case, $\cosh(t_\text{M} \tau) \approx \sinh(t_\text{M} \tau)$, and the impurity state approaches the ground state $|\psi_1\rangle$, which has an opposite parity from the initial one $|\psi_2\rangle$, before the next lead-dot tunneling.
The RG flow is then distinct from that of a regular DRLM.

\section{Low-temperature RG flow equations}

In the section we provide detailed derivations of the low-temperature RG flow equations, Eq.~(3) of the main text.
Briefly, within the Coulomb-gas RG formalizm, one deals with the partition function~\cite{SKaneFisherPRB92}
\begin{equation}
    Z = \sum_{\left\{q_j\right\}} \sum_{\left\{d_j \right\}} \prod_j t_j \Big\langle \Pi_j e^{i\frac{1}{\sqrt{2}} [q_j \phi_c (\tau_j) + d_j \phi_f (\tau_j) ] }\Big\rangle_{H_\text{leads}} \Big\langle 
\Pi_j D (\varepsilon_j,\tau_j) \Big\rangle_{H_\text{impurity}}
\end{equation}
for a two-lead system, where $t_j = V_L, V_R$ refers to the amplitude of the $j$th tunneling event that occurs at time $\tau_j$. This tunneling event contains the impurity and lead operators $D (\varepsilon_j,\tau_j)$ and $\exp\left\{i\frac{1}{\sqrt{2}} [q_j \phi_c (\tau_j) + d_j \phi_f (\tau_j) ]\right\}$, respectively.
Here $q_j$ and $d_j$ reflect the change of total lead charge ($N_L + N_R$ of the main text), and the charge difference in two leads ($N_L - N_R$ of the main text), respectively: they are charges considered in a regular DRLM~\cite{SDongHuaixiuPRB14}.
The dot operator $D $ equals $d^\dagger (\tau_i)$ and $d (\tau_i)$ if $\varepsilon = \pm 1$, respectively.

\begin{figure}
  \includegraphics[width = 1 \linewidth]{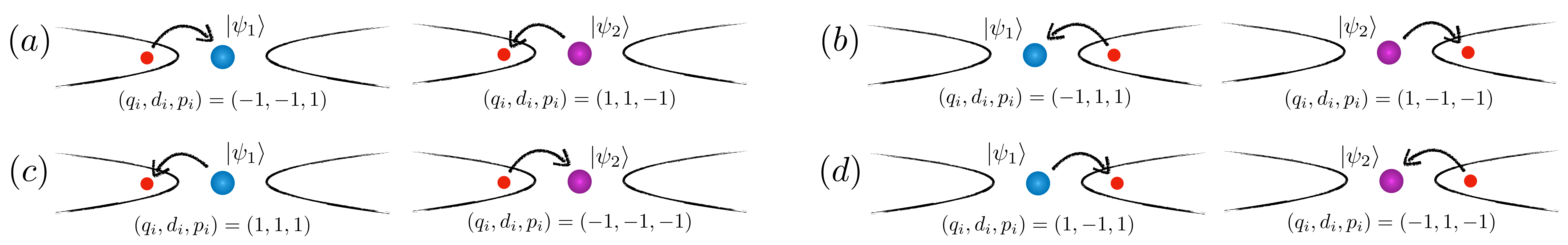}
  \caption{Four possible lead-impurity tunneling processes, of the $i$th tunneling. The cyan and purple dot refer to two different impurity ground states $|\psi_1\rangle$ and $|\psi_2\rangle$, respectively. The red small dot refers to a tunneling electron. In each figure, the process at the right side is the inverse process of that at the left side. Charges $q_i$ and $d_i$ refer to the change of the total charge number, and the charge number differences in two leads, respectively.
  The third charge $p_i$ refers to the change of parity of the impurity part.
  }
  \label{fig:low_t_processes}
\end{figure}

For the low-temperature situation, $\sinh (\tau  t_\text{M}) \approx \cosh (\tau  t_\text{M}) \approx \exp(|\tau t_\text{M}|)$.
In addition, in this regime, the parity of the impurity part (instead of the dot occupation number of the high) alternates in time.
We thus introduce another charge number, the parity charge $p_j$, to indicate the change of parity of the impurity part, before the $j$th tunneling event.
The partition function then becomes
\begin{equation}
    \mathcal{Z} \!= \!\sum_N \sum'_{\left\{q_j\right\}} \sum'_{\left\{ d_j \right\}} \sum'_{\left\{ p_j \right\}} \prod_{j = 1}^N t_j \int _0^{\beta - \tau_c} \! d\tau_{N} \! \cdot\cdot\cdot \!\int_0^{\tau_{j+1} - \tau_c} \! d\tau_{j} \! \int_0^{\tau_{j} - \tau_c} \!d\tau_{j-1}\! \cdot \cdot \cdot \int_0^{\tau_2 - \tau_c} \!d\tau_1  e^{-\sum_{i<j} \mathcal{V}_{ij}} \prod_{j= 1}^{N-1} e^{|\tau_j - \tau_{j+1}| t_\text{M}}
    \label{eq:partition_low_t}
\end{equation}
where prime indicates $\sum_j q_j = \sum_j d_j = \sum_j p_j = 0$, and the ``potential interaction'' equals
\begin{equation}
   \mathcal{V}_{ij} = \frac{1+r}{2} \left[ \frac{1}{1+r} q_i q_j + d_i d_j + K_\text{parity} p_i p_j + K_\text{asy} (p_i d_j + p_j d_i) + \frac{1}{1 + r} K_\text{asy}' ( 
p_i q_j + p_j q_i ) \right] \ln (\tau_i - \tau_j)/\tau_c,
\end{equation}
where $K_\text{parity}$ is the scaling factor of parity charge. Its initial value equals zero, since the lead fermion, after bosonization, contains only the common and different fields.
Due to the alternating parity, another asymmetric scaling $K_\text{asy}'$ is introduced, refering to the asymmetric scaling of the charge sector. It also has the initial value zero.

\begin{figure}
  \includegraphics[width = 0.5 \linewidth]{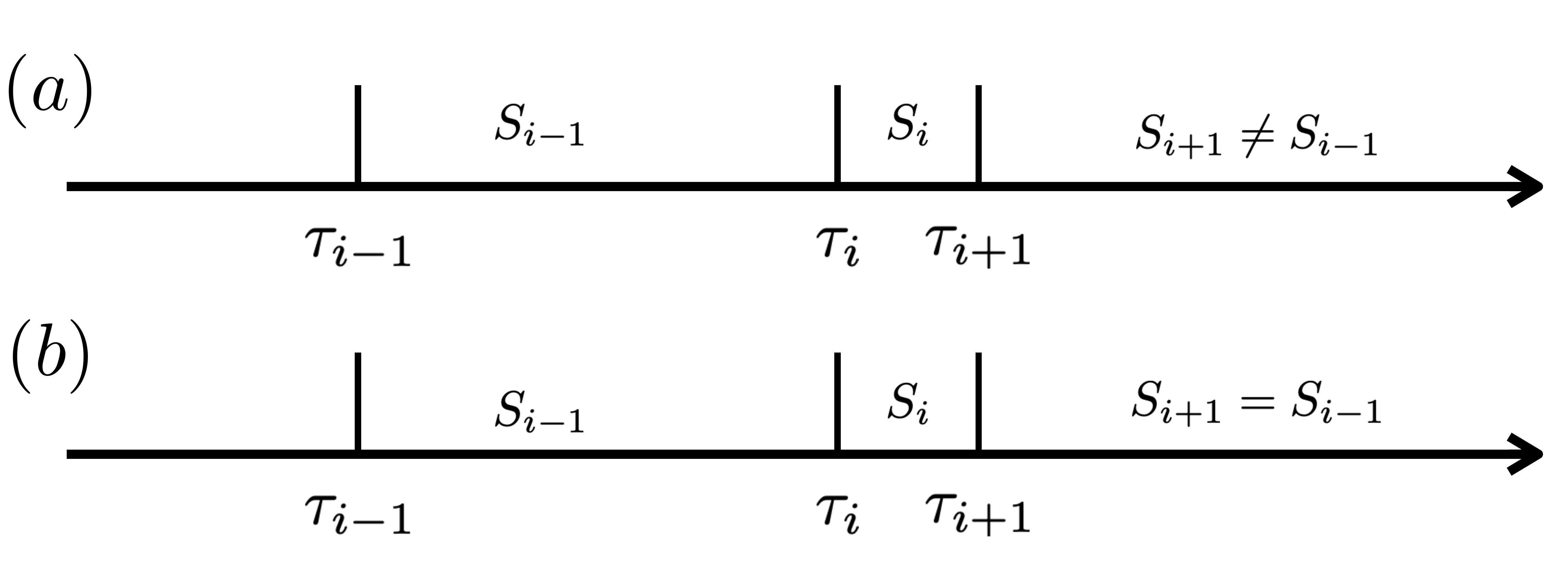}
  \caption{Two different pairs to deal with in Coulomb gas RG.
  (a) Of the non-neutral pair situation, the system state $S_{i-1}$ before the close kink pairs (that occur at $\tau_i$ and $\tau_{i+1}$, respectively) are different from that $S_{i+1}$ after the pairs. One needs to combine non-neutral kinks at $\tau_i$ and $\tau_{i+1}$ into a single one. (b) When the pre and post-kink-pair states equal $S_{i-1} = S_{i+1}$, kinks at both $\tau_i$ and $\tau_{i+1}$ should be removed.}
  \label{fig:cg}
\end{figure}

Following Eq.~\eqref{eq:partition_low_t}, two neighbouring kinks are not allowed to occur within the time cutoff $\tau_c$.
At the starting of each RG step, the cutoff increases to $\tau_c + d\tau_c$. Neighbouring kinks then have the chance to occur within the new cutoff, i.e., $\tau_c < |\tau_i - \tau_{i+1}| < \tau_c + d\tau_c$.
There are two different close kink pairs, as shown in Fig.~\ref{fig:cg}. In Fig.~\ref{fig:cg}(a), the state $S_{i-1}$ and $S_{i+1}$ are different, called as the non-neutral pair. Of this case, we can simply combine two kinks at $\tau_i$ and $\tau_{i+1}$.
Of the second case, in Fig.~\ref{fig:cg}(b), these two states equal, known as the neutral pair. Of this case, both kinks need to be integrated out.
In this work the BKT QPT occurs when transmissions are small.
We thus ignore the non-neutral pairs.
The inclusion of neutral pairs changes only the critical value of the coupling constant $V_C$, but not the qualitative feature.

The integral over neutral pair leads to the extra potential interaction $\delta \mathcal{V}_{ij}$
\begin{equation}
 \delta \mathcal{V}_{ij} \!= \! \sum_{\alpha} t_i^2 \frac{1\!+\!r}{2} \left[ \frac{1}{1\!+\!r} q_i q_j \! + \! d_i d_j \! +\! K_\text{parity} p_i p_j \!+\! K_\text{asy} (p_i d_j \!+\! p_j d_i) \!+\! \frac{1}{1 \!+\! r} K_\text{asy}' ( 
p_i q_j \!+\! p_j q_i ) \right] \! \tau_c \partial_\tau \ln (\tau = \tau_j)/\tau_c,
\end{equation}
after summation over all possible kink choices shown in Fig.~\ref{fig:low_t_processes}.
Notice that since the parity charge alternates in time, $p = p_i$ is enforced, for all possible neutral pairs.
One can then express other two charges of the neutral pair in terms of the parity charge.

After summing over all possible kinks, we arrive at extra potential generated from the neutral pair
\begin{equation}
    \delta \mathcal{V}_j = - 2 (1 + r) \tau_c (V_L^2 + V_R^2) \left[  K_\text{parity} p p_j + K_\text{asy} p d_j + K_\text{asy}' p q_j  \right] d\tau_c \ln (\tau_i - \tau).
\end{equation}
The extra contribution above can be combined with other kinks, leading to the renormalization of scaling parameters
\begin{equation}
\frac{dK_\text{parity}}{d l} = -4 \tau_c^2 (V_L^2 + V_R^2) K_\text{parity},\  \frac{dK_\text{asy}}{d l} = -2 \tau_c^2 (V_L^2 + V_R^2) K_\text{asy},\ \frac{dK_\text{asy}'}{d l} = -2 \tau_c^2 (V_L^2 + V_R^2)  K_\text{asy}'.
\label{eq:low-t-flow}
\end{equation}
From Eq.~\eqref{eq:low-t-flow}, we see that $K_\text{parity}$ and $K_\text{asy}'$, with vanishing initial values, are both fixed at zero: the parity charge will not introduce extra asymmetric scaling parameters.
In addition, although $K_\text{asy}$ is finite at the end of the high-temperature flow, it gradually decreases to zero in the low-temperature flow.
The asymmetry is thus irrelevant in the low-temperature regime.

\section{RG flow of a finite dot energy $\epsilon_d$}

In the main text, we take $\epsilon_d = 0$, to focus on the effect of a left-right asymmetry $V_L - V_R$.
In this section, we take a finite $\epsilon_d$, to study whether the topological-parity degeneracy can also ``restore'' the detuning, or ``particle-hole'' asymmetry at the dot.

For the high-temperature situation, Majorana-coupling $t_\text{M}$ is negligible, and RG flow mimics that of a regular DRLM. RG flow equations with a finite $\epsilon_d$ then become (after reducing the spinful equations of Ref.~\cite{SSchillerIngersentRG97} to the spinless case)
\begin{equation}
\begin{aligned}
& \frac{d V_{L,R}}{dl} = V_{L,R} \left[ 1 \! - \! \frac{1+r}{4} (1\! +\! K_\text{charge} \!\pm \! 2 K_\text{asy}) \right],\\
& \frac{d K_\text{charge}}{dl} \! = \! -2 \tau_c^2 [K_\text{charge} (V_L^2 \! + \! V_R^2) \!+\! K_\text{asy} (V_L^2 \!-\! V_R^2) ] \left(e^{\epsilon_d \tau_c} + e^{-\epsilon_d \tau_c} \right), \\
& \frac{d K_\text{asy}}{dl} \!=\! - \tau_c^2 [K_\text{asy} (V_L^2 \!+\! V_R^2) \!+\! (V_L^2 \!-\! V_R^2) ] \left(e^{\epsilon_d \tau_c} + e^{-\epsilon_d \tau_c} \right),\\
& \frac{d\epsilon_d}{dl} = \epsilon_d + (V_L^2 + V_R^2) \tau_c\left(e^{-\epsilon_d \tau_c} - e^{\epsilon_d \tau_c} \right) .
\end{aligned}
\label{eq:finite-epsilon-high}
\end{equation}
Clearly a finite $\epsilon_d$ is RG relevant, and increases when decreasing the temperature.
When $t_\text{M} = 0$, the lead-dot tunnelings are then turned off at zero temperature, where the dot is either fully unoccupied (when $\epsilon_d < 0$), or empty (when $\epsilon_d > 0$).
In addition, a finite $\epsilon_d$, either positive or negative, accelerates the flow of scaling factors $K_\text{asy}$ and $K_\text{charge}$.
Following Eq.~\eqref{eq:finite-epsilon-high}, in the high-temperature regime, a finite dot detuning $\epsilon_d$ increases the system asymmetry, and tend to sabotage the BKT transition.

In the low-temperature regime, RG flow features above are however greatly modified. Indeed, in this regime, the dot operator has the following influence on the ground states
\begin{equation}
\begin{aligned}
  &  d |\psi_1\rangle  = -\frac{i}{2} ( \mathcal{A} |\psi_2\rangle + 
\mathcal{B} |\psi_4\rangle ),\ \  & d^\dagger |\psi_1 \rangle  = \frac{i}{2} (\mathcal{B} |\psi_2\rangle - \mathcal{A} |\psi_4\rangle),\\
& d |\psi_2\rangle  = -\frac{i}{2} ( \mathcal{A} |\psi_1\rangle + 
\mathcal{B} |\psi_3\rangle ),\ \  & d^\dagger |\psi_1 \rangle  = \frac{i}{2} (\mathcal{B} |\psi_1\rangle - \mathcal{A} |\psi_3\rangle),
\end{aligned}
\end{equation}
with two $\epsilon_d$-dependent positive parameters
\begin{equation}
    \mathcal{A} = \sqrt{1 + \frac{\epsilon_d}{\sqrt{\epsilon_d^2 + 4t_\text{M}}}}, \ \ \mathcal{B} = \sqrt{\frac{4t_\text{M}^2}{4 t_\text{M}^2 + \epsilon_d (\epsilon_d + \sqrt{\epsilon_d^2 + 4 t_\text{M}^2})}}.
\end{equation}
As the consequence, we obtain the low-temperature RG equations
\begin{equation}
\begin{aligned}
    & \frac{d V_{L,R}}{dl} \!=\! V_{L,R} \left[ 1 \! - \! \frac{1+r}{4} (1\! +\! K_\text{charge} \!\pm \! 2 K_\text{asy}) \right], \frac{dK_\text{parity}}{d l} = -4 \mathcal{A} \mathcal{B} \tau_c^2 (V_L^2 + V_R^2) K_\text{parity}, \\
    & \frac{dK_\text{asy}}{d l} = -2 \mathcal{A} \mathcal{B} \tau_c^2 (V_L^2 + V_R^2) K_\text{asy},\ \frac{dK_\text{asy}'}{d l} = -2 \mathcal{A} \mathcal{B} \tau_c^2 (V_L^2 + V_R^2) K_\text{asy}',\ \frac{dK_\text{charge}}{d l} = 0,
\end{aligned}
\label{eq:finite-epsilond-low}
\end{equation}
where the $\epsilon_d$-dependent factors $\mathcal{A}$ and $ \mathcal{B}$ influence the flow of all scaling parameters.
However, since $\mathcal{A} \mathcal{B} > 0$, the flow patterns of all scaling parameters qualitatively agree with that of the fine-tuned situation, i.e., Eq.~\eqref{eq:low-t-flow}, and Eq.~(3) of the main text.
A finite $\epsilon_d$ simply modifies the prefactors of RG flow equations. These modifications change only the critical value of $V_L$, $V_R$ at the boundary of the transition. The qualitative feature of the phase diagram however remains invariant.

\section{Coupling the DRLM dot to an Andreev bound state}

In the main text, we have shown that by coupling the dot of a DRLM to an MZM, the BKT QPT hosted by the DRLM will be greatly modified.
It is also illustrated and highlighted that the modification of the BKT phase diagram is due to the topological degeneracy provided by the Majorana.
To better understand the importance of the topological degeneracy, we consider another example where the dot instead couples to an Andreev bound state (ABS, with the state operator $d_\text{ABS}$), with has instead a trivial local degeneracy. Of this case, the impurity part has the Hamiltonian
\begin{equation}
    H_\text{impurity} = \epsilon_\text{dot} d^\dagger d + \epsilon_\text{ABS} d_\text{ABS}^\dagger d_\text{ABS} + t_1 ( d_\text{dot}^\dagger d_\text{ABS} + d_\text{ABS}^\dagger d_\text{dot} ) + t_2 ( d_\text{dot}^\dagger d^\dagger_\text{ABS} + d_\text{ABS} d_\text{dot} ),
    \label{eq:abs_tunneling}
\end{equation}
where $t_1 = t_2$ for an MZM. For simplicity we have assumed $t_1$ and $t_2$ to be both real numbers.
The impurity Hamiltonian has four eigenvalues,
\begin{equation}
\begin{aligned}
    \epsilon_1 & = \frac{1}{2} \left[ \epsilon_\text{dot} + \epsilon_\text{ABS} - \sqrt{(\epsilon_\text{dot} + \epsilon_\text{ABS})^2 + 4 t_1^2 } \right], \ \ \epsilon_2 = \frac{1}{2} \left[ \epsilon_\text{dot} + \epsilon_\text{ABS} - \sqrt{(\epsilon_\text{dot} - \epsilon_\text{ABS})^2 + 4 t_2^2 } \right],\\
    \epsilon_3 & = \frac{1}{2} \left[ \epsilon_\text{dot} + \epsilon_\text{ABS} + \sqrt{(\epsilon_\text{dot} + \epsilon_\text{ABS})^2 + 4 t_1^2 } \right], \ \ \epsilon_4 = \frac{1}{2} \left[ \epsilon_\text{dot} + \epsilon_\text{ABS} + \sqrt{(\epsilon_\text{dot} - \epsilon_\text{ABS})^2 + 4 t_2^2 } \right].
\end{aligned}
\end{equation}
Apparently, energies $\epsilon_1$ and $\epsilon_2$ are candidates of the ground-state energy. If either $\epsilon_\text{dot}$ or $\epsilon_\text{ABS}$ (corresponding to an accidentally fine-tuned ABS) is zero, $\epsilon_1 \neq \epsilon_2$ and the dot-ABS state reduces an effectively detuned dot: of this case, the BKT QPT does not exist, in great contrast to the enhanced BKT phase diagram of the MZM situation.

\begin{figure}
  \includegraphics[width = 0.8 \linewidth]{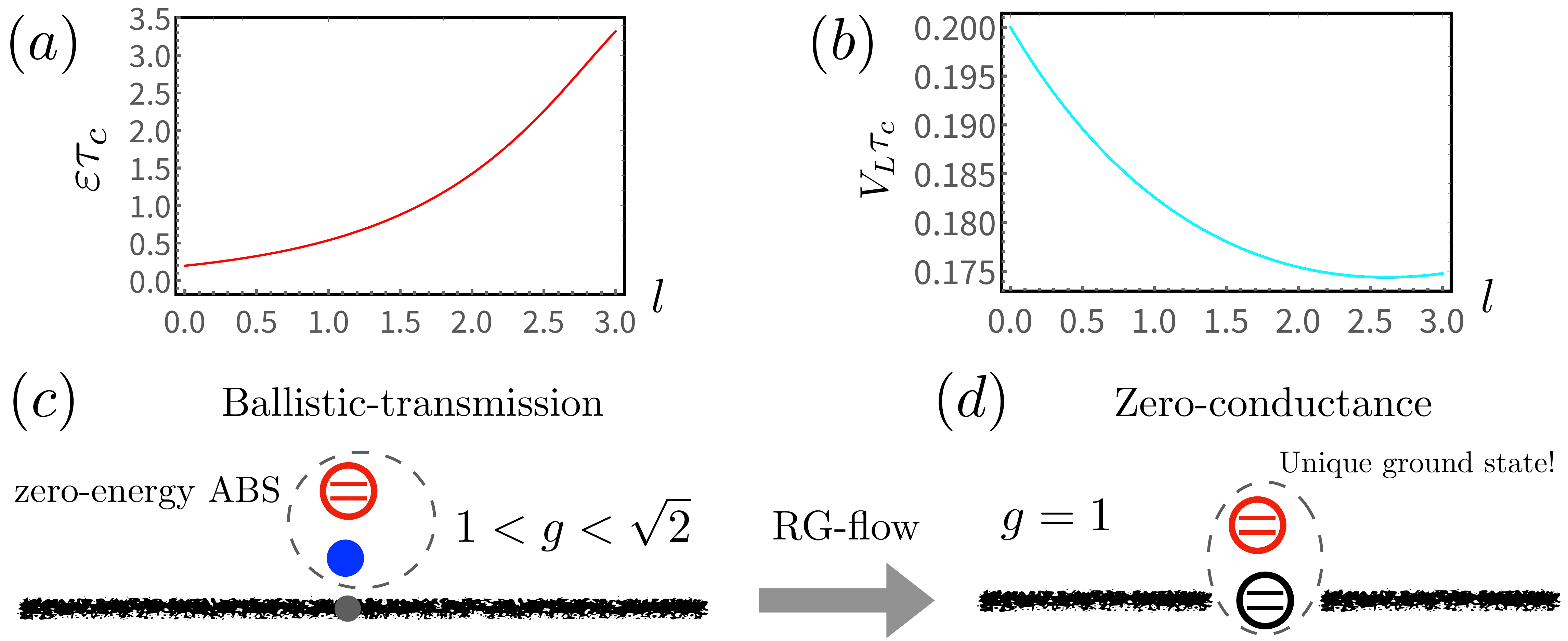}
  \caption{RG flows [(a) and (b)] and the corresponding g-theorem analysis [(c) and (d)] of the situation where the resonant level couples to a zero-energy Andreev bound state (the red circle). (a) and (b) are the RG flows of $\varepsilon \equiv t_1 - t_2$ [see Eq.~\eqref{eq:abs_tunneling}], when $\varepsilon \tau_c = 0.2$, and $V_L \tau_c = V_R \tau_c = 0.2$ (here we assume the left-right symmetry to focus on the a finite $\varepsilon$). The value of $\varepsilon$ has become rather significant (a), while the lead-dot tunneling remains small (b). When coupling to a zero-energy Andreev bound state, the impurity part has a unique ground state (d). The zero-conductance phase is thus more preferred at zero energy, in great contrast to that of the Majorana situation.}
  \label{fig:abs_situations}
\end{figure}

For simplicity, we can consider the special situation $\epsilon_\text{dot} = \epsilon_\text{ABS} = 0$ where both the dot and Andreev bound state energies equal zero.
We take this example since a zero-energy Andreev bound state is known as among the factors that greatly undermine the reliability of potential Majorana signals in non-interacting systems.
Of this case, the impurity part has eigenstates and corresponding eigen-energies
\begin{equation}
\begin{aligned}
& |\phi_1\rangle = \frac{-|1,0\rangle + | 0,1\rangle}{\sqrt{2}} ,\ \varepsilon_1 = -t_1; \ \ \ \ \ |\phi_2 \rangle = \frac{-|1,1\rangle + | 0,0\rangle}{\sqrt{2}},\ \varepsilon_2 = -t_2;\\
& |\phi_3\rangle = \frac{|1,0\rangle + | 0,1\rangle}{\sqrt{2}} ,\ \varepsilon_3 = t_1; \ \ \ \ \ \ \ \  \ \ \ |\phi_4 \rangle = \frac{|1,1\rangle + | 0,0\rangle}{\sqrt{2}},\ \varepsilon_4 = t_2.
\end{aligned}
\label{eq:abs_eigen}
\end{equation} 
Without loss of generality, we assume $t_1>t_2 > 0$. States with two lowest energies then have the energy difference $\varepsilon \equiv t_1 - t_2$. $\varepsilon$ obeys the RG flow
\begin{equation}
    \frac{d}{dl}\varepsilon = \varepsilon + \frac{\tau_c (V_L^2 + V_R^2)}{4} (e^{- \varepsilon\tau_c} - e^{ \varepsilon\tau_c}), 
    \label{eq:abs_energy}
\end{equation}
which grows when the energy decreases.
For a more intuitive understanding, we plot the RG flow of $\varepsilon$ and $V_L$ in Fig.~\ref{fig:abs_situations}, where $\varepsilon \tau_c = 0.2$, and we have assumed the left-right symmetry $V_L \tau_c = V_R \tau_c = 0.2$, to focus on the influence of the finiteness of $\varepsilon$.
Clearly, $\varepsilon$ has become rather significant [in Fig.~\ref{fig:abs_situations}(a)], while $V_L$ remains rather small [in Fig.~\ref{fig:abs_situations}(b)].
As the consequence, the value of $\varepsilon$, which might be small initially, becomes significant at low energies, thus suppressing tunneling through the resonant level.
Physically, the relevance of $\varepsilon$ can be understood via the corresponding g-theorem in Figs.~\ref{fig:abs_situations}(c) and \ref{fig:abs_situations}(d).
Briefly, when $t_1 \neq t_2$, the zero-energy Andreev bound state and the fine-tuned dot combine into a unique ground state. The zero-conductance phase [Fig.~\ref{fig:abs_situations}(d)] then has a smaller degeneracy ($g = 1$) than that of the ballistic-transmission phase [($ 1 < g < \sqrt{2}$), Fig.~\ref{fig:abs_situations}(c)]. The zero-conductance phase is thus the preferred phase at zero temperature, which greatly contrasts that of the Majorana situation.

Specially, these two energies only equal when $\epsilon_\text{dot} \epsilon_\text{ABS} + t_1^2 = t_2^2$, leading to an accidental degenerate state.
However, this accidental degeneracy will neither simply the observation of the BKT transition, nor undermine the credibility of the extended ballistic-transmission phase as an MZM evidence.
Indeed, on one hand, although this accidental degeneracy can relax the left-right symmetry requirement to observe the BKT transition, its realization introduces extra complexities.
On the other hand, in real experiments, the tuning of $V_L$ and $V_R$, with which the phase diagram is detected, will effectively tune the dot energy $\epsilon_\text{dot}$ (which may also influence $\epsilon_\text{ABS}$). Indeed, the cross-talk between different gates is an experimentally common phenomenon (see, e.g., Ref.~\cite{SMebrahtu12}). As the consequence, we expect that the accidental degenerate situation only leads to scattered full-transmission points in phase space, being distinct from the real MZM situation.

Following analysis above, we clearly see that the Majorana-induced modification of the low-temperature RG flow, physically explained in Fig.~3, does require the presence of a topological degeneracy.
A fine-tuned trivial ABS ($\epsilon_\text{ABS} = 0$) instead leads to always the isolated impurity phase.

Before ending this section, we stress that following discussions above, the BKT-based Majorana detection is also capable to work on systems where the hybrid nanowire contains multiple conductance channels.
Briefly, with multiple conductance channels, the hybrid nanowire either hosts one Majorana and multiple Andreev bound states, or hosts only multiple Andreev bound states. Notice that even numbers of Majorana, if hosted by the nanowire, normally recombine into one or multiple Andreev bound states.
Following analysis in this section, these Andreev bound states, if any, will not stabilize the BKT transition.
Indeed, even with multiple conductance channels, only the Majorana, if hosted, can stabilize the BKT at zero temperature (even if accompanied by Andreev bound states). The corresponding phase diagram is thus in great contrast to that when the hybrid nanowire hosts only Andreev bound states, where the BKT phase diagram is unstabilized.
Two dot-coupled leads, on the other hand, can contain only one conductance channel, as has been experimentally realized, by e.g.,~\cite{SMebrahtu12}.
One thus does not need to worry about multiple conductance channels in the dot-coupled leads.

\section{Majorana identification with BKT transition}

As stated in the main text, we anticipate that our protocol, which detects the irrational degeneracy of a Majorana zero mode, can provide another piece of evidence to further confirm the plausible Majorana signals. Below we provide a protocol to specify our method.

1.~\emph{Evaluating the dissipation amplitude of the dissipative resonant level} --- This evaluation can be realized when the dissipative resonant level remains in the weak-tunneling regime, which does not require a fine-tuned dot, or coupling to the superconducting island (i.e., $t_\text{M} = 0$).
More specifically, within the weak-tunneling limit, tunneling through the dissipative resonant level displays the scaling feature $G \propto \max (V,T)^{2r}$ (at low energies), where $V$ and $T$ refer to the bias and temperature, respectively, and $r = R e^2/h$ is the dimensionless dissipation amplitude.
The dissipation is qualified for further operation, if $2< r < 3$ following scaling analysis. Actually, $r \sim 2.5$ is preferred, to observe both phases, under a reasonable temperature.

2.~\emph{Obtain the phase diagram with a finite $t_\text{M}$}.
As the second step, we couple the dissipative resonant level model to the hybrid lead on top of the superconducting island (which has shown the potential to be within the topological regime),
and study the zero-bias conductance (through the dissipative resonant level) as a function of three gate voltages, that control the dot-left lead ($V_\text{LG}$), dot-right lead ($V_\text{RG}$), and dot energy level couplings ($V_\text{SG}$).
If the hybrid nanowire contains a Majorana, we anticipate to observe two finite-size phase spaces: a continuous space with a zero-bias conductance peak, and that where the conductance abruptly decreases to zero.
Notice that a quantized zero-bias conductance peak is not required in this step, as a strong dissipation will reduce the temperature at which the conductance saturates to the quantized value.
A zero-bias conductance already suffices to tell trivial from non-trivial phases.

3.~\emph{Further confirm the Majorana presence by decreasing the temperature}.
Given step 2 fulfilled, we can further confirm the Majorana presence by visiting the temperature-dependence of the zero-bias conductance peak.
For a real Majorana-hosted island, two features are anticipated when decreasing the temperature:
(i) A decreasing width of the boundary between two phases observed in step 2 (i.e., the phase space where zero-bias conductance changes abruptly), and (ii) An increasing zero-bias conductance for the phase space with a zero-bias conductance peak, and an decreasing conductance if outside of this phase.

4.~\emph{Confirm the Majorana influence}.
As the last step, one can turn off the coupling between the Majorana zero mode and the resonant level (dot).
By doing so, the ballistic-transmission phase once again requires fine-tuning all gate voltages.
The corresponding phase space would thus become greatly modified.
Due to the cross-talk between different gates, the ballistic-transmission phase will possibly become discrete, or even hardly capturable at low temperatures.


\begin{thebibliography}{85}
\bibitem [{Foo()}]{Footnote}%
  \BibitemOpen
  \href@noop {} {}\bibinfo {note} {The QPT is not of BKT-type when $r<2$~\cite{Mebrahtu12}. When $r>3$, the dot is always isolated.}\BibitemShut {Stop}%
\end{thebibliography}
\end{document}